\documentclass[11pt,a4paper]{article}
\pdfoutput=1   
\usepackage{amsmath,amsfonts,amssymb,amsbsy}
\usepackage{mathrsfs,latexsym}
\usepackage{graphicx}
\usepackage{color}
\usepackage{booktabs}
\usepackage{multirow}
\usepackage{multicol}
\usepackage[]{alpha}
\hypersetup{%
   pdftitle = {Decay constants of B-mesons from non-perturbative HQET with two light dynamical quarks},
   pdfauthor = {F.Bernardoni, B.Blossier, J.Bulava, M.DellaMorte, P.Fritzsch, N.Garron, A.Gerardin, J.Heitger, G.vonHippel, H.Simma, R.Sommer},
   pdfkeywords = {Lattice QCD, Heavy Quark Effective Theory}
}%
\usepackage{macros}
\graphicspath{{plots/}}

\begin{document}

\preprintno{%
DESY 14-048\\
HU-EP-14/14\\
IFIC/14-24\\
CP3-Origins-2014-011 DNRF90\\
DIAS-2014-11\\
LPT-Orsay/14-19\\
MITP/14-026\\
MS-TP-14-18\\
SFB/CPP-14-20\\
TCD 14--03\\
}

\title{%
Decay constants of B-mesons from non-perturbative HQET with two light dynamical quarks
}

\collaboration{\includegraphics[width=2.8cm]{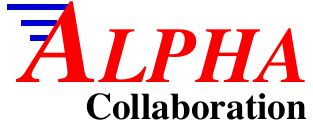}}

\author[desy]{Fabio~Bernardoni}
\author[fra]{Beno\^it~Blossier}
\author[trin]{John~Bulava}
\author[ode,esp]{Michele~Della~Morte}
\author[hu]{Patrick~Fritzsch}
\author[trin]{Nicolas~Garron}
\author[fra]{Antoine~G\'erardin}
\author[wwu]{Jochen~Heitger}
\author[prism]{Georg~von~Hippel}
\author[desy]{Hubert~Simma}
\author[desy]{Rainer~Sommer}

\address[desy]{NIC @ DESY, Platanenallee~6, 15738~Zeuthen, Germany}
\address[fra]{Laboratoire~de~Physique~Th\'eorique, Universit\'e~Paris~XI,  91405~Orsay~Cedex, France}
\address[trin]{School~of~Mathematics, Trinity~College, Dublin~2, Ireland}
\address[ode]{CP$^3$-Origins \& Danish IAS, University of Southern Denmark, Campusvej 55, 5230 Odense M, Denmark}
\address[esp]{IFIC and CSIC, Calle Catedr\'atico Jos\'e Beltran 2, 46980~Paterna, Valencia, Spain}
\address[hu]{Institut~f\"ur~Physik, Humboldt-Universit\"at~zu~Berlin, Newtonstr.~15, 12489~Berlin, Germany}
\address[wwu]{Institut~f\"ur~Theoretische~Physik, Universit\"at~M\"unster, Wilhelm-Klemm-Str.~9, 48149~M\"unster, Germany}
\address[prism]{PRISMA~Cluster~of~Excellence, Institut~f{\"u}r~Kernphysik, University~of~Mainz, Becherweg~45, 55099~Mainz, Germany}

\begin{abstract}
We present a computation of B-meson decay constants from lattice QCD
simulations within the framework of Heavy Quark Effective Theory for the
b-quark. The next-to-leading order corrections in the HQET expansion are
included non-perturbatively. Based on $\Nf=2$ gauge field ensembles, covering
three lattice spacings $a\approx(0.08-0.05)\,\fm$ and pion masses down to
$190\,\MeV$, a variational method for extracting hadronic matrix elements is
used to keep systematic errors under control.  In addition we perform a careful
autocorrelation analysis in the extrapolation to the continuum and to the
physical pion mass limits.  Our final results read $\fB=186(13)\,\MeV$,
$\fBs=224(14)\,\MeV$ and $\fBs/\fB=1.203(65)$.  A comparison with other results
in the literature does not reveal a dependence on the number of dynamical
quarks, and effects from truncating HQET appear to be negligible.
\end{abstract}


\maketitle

\section{Introduction}\label{sec:intro}

In the ongoing quest for new effects in high-energy particle physics, flavour
physics provides information complementary to that from the direct searches
performed at ATLAS and CMS.  Indeed, low-energy processes and rare events can
be sensitive probes of New Physics, in particular when they are mediated by
virtual loops, in which non-Standard Model particles can circulate, or when
they involve new couplings occurring at tree-level.  However, any analysis of
experimental data in the quark sector depends on theoretical inputs, such as
hadron decay constants, that encode the long-distance dynamics of QCD, which
cannot be reliably estimated in perturbation theory.

In this regard, B-physics is an emblematic case.  For example, it is crucial
to understand the origin of the current discrepancy in the Cabibbo--Kobayashi--Maskawa
matrix element $V_{\rm ub}$ measured through the exclusive processes 
$B \to \tau \nu$
\cite{Lees:2012ju, Adachi:2012mm}
and  $B \to \pi \ell \nu$ 
\cite{delAmoSanchez:2010zd, Ha:2010rf}, 
where the latter makes use of the $B \to \pi$ form factors computed on the
lattice. Is it due to an experimental problem, or due
to new physics such as the presence of a new, right-handed, 
tree-level coupling to a charged Higgs boson in the $B$ leptonic decay
\cite{Hou:1992sy},
or due to a severe underestimate of the uncertainty on the decay constant $\fB$
governing that decay.  For comparison, the recent measurements of 
$\mcB(B_{\rm s} \to \mu^+ \mu^-)$ at LHC
\cite{Aaij:2013aka,Chatrchyan:2013bka}
are in excellent agreement with the Standard Model prediction
\cite{Buras:2012ru, Buras:2013uqa},
where the latter depends on the decay constant $\fBs$, whose estimate is
dominated by lattice results.

The methods that have been used to estimate $\fB$ and $\fBs$ include
applications of quark models, as discussed in
\cite{Morenas:1997rx, Ebert:2006hj, Badalian:2007km}
and references therein, and QCD sum rules in the analysis of two-point B-meson
correlators
\cite{ Jamin:2001fw, Lucha:2010ea, Narison:2012mz, Baker:2013mwa}.
Several strategies have been proposed to determine $\fB$ and $\fBs$ from first
principles using lattice field theory, including the extrapolation of
simulation results obtained in the region between the charm quark mass $\mc$
and a mass $\sim 3 \mc$ to the physical b-quark mass $\mb$
\cite{Dimopoulos:2011gx,Carrasco:2013zta},
simulations of relativistic b-quarks using an action tuned so as to minimize
discretization errors
\cite{McNeile:2011ng, Bazavov:2011aa},
and the use of Non-Relativistic QCD 
\cite{Na:2012kp, Dowdall:2013tga}.
We here use Heavy Quark Effective Theory (HQET)
\cite{stat:eichhill1,stat:symm1,stat:symm3,Eichten:1990vp},
regularized on the lattice with the parameters of HQET 
determined by a non-perturbative matching to QCD 
\cite{Heitger:2003nj,DellaMorte:2006cb,Blossier:2010jk}.
The virtue of this approach is that perturbative errors are absent and the
continuum limit exists.  The matching at order $\Or(\minvh)$ has been performed
in the $\Nf=2$ theory
\cite{Blossier:2012qu};
this is the first step required, for example, in order to compute the b-quark
mass
\cite{Bernardoni:2013xba},
which we use in this \paperorletter~to extract $\fB$, $\fBs$ and $\fBs/\fB$
from our simulations.  In Section \ref{s:method}, we review the methods
employed, before presenting the results in Section~\ref{sec:extrapolation}.
Section \ref{sec:conclusions} contains our conclusions.

\section{Methodology}\label{s:method}

\subsection{HQET on the lattice}\label{s:lathqet}

\noindent Heavy Quark Effective Theory regularized on the lattice is a
well-defined approach to B-physics.  It is based on an expansion in powers of
${\minvh}$ of QCD correlation functions around the limit $\minvh \to 0$.  The
continuum limit can be taken order by order in the expansion, since it only
requires correlation functions computed in the static theory, which is
non-perturbatively renormalizable.
Applying the strategy previously discussed in
\cite{lat02:rainer,Heitger:2003nj,Blossier:2010jk}
and employed to measure $\fBs$ in the quenched approximation
\cite{Blossier:2010mk},
the HQET action and the time component of the axial current expanded to
$\Or(\minvh)$ read
{\setlength{\jot}{7pt}
\begin{align}         \label{e:hqetaction}
  \Shqet 
      &= a^4{\sum}_x \big\{ \Lstat(x) - \omega_{\rm kin} \Okin(x) - \omega_{\rm spin} \Ospin(x) \big\} \,,  \\ \label{e:lstat}
  \Lstat(x) 
      &= \heavyb(x)\,  D_0                  \,\heavy(x)\,, \\ \label{e:ofirst}
  \Okin(x) 
      &= \heavyb(x)\, \vecD^2               \,\heavy(x)\,, \\
  \Ospin(x) 
      &= \heavyb(x)\, \vecsigma \cdot \vecB \,\heavy(x)\,,
        \label{}
\end{align}}%
and {\setlength{\jot}{7pt}%
\begin{align}      \label{e:hqetcurrent}
  \Ahqet(x) 
     &= \zahqet \Big[ \Astat(x) + {\sum}_{i=1}^2 \cah{i}\Azero{(i)}(x) \Big] \;, \\ \label{e:impcur}
  \Azero{(1)}(x) 
     &= \psibar_{q}\, \tfrac{1}{2}\gamma_5\gamma_i(\nabS{i} - \nabSleft{i}) \, \heavy(x)  \;, \\
  \Azero{(2)}(x) 
     &= \psibar_{q}\, \tfrac{1}{2}\gamma_5\gamma_i(\nabS{i} + \nabSleft{i}) \, \heavy(x)  \;, \quad
  \Astati(x) = \psibar_{q}(x)\, \gamma_i\gamma_5 \,\heavy(x)        \,.
\end{align}
}%
We use the labels $\rm h$ to denote the heavy (static) quark field appearing in
the HQET Lagrangian, and $q={\rm u/d}, {\rm s}$ for the light and strange quark
channels, respectively.  The normalization is such that the classical values of
the coefficients are $\omegakin=\omegaspin=-\cah{i}=1/(2\mh)$.  A bare mass
$\mhbare$ has to be added to the energy levels computed with this Lagrangian in
order to obtain the QCD ones. At the classical level it is $\mh$, but in the
quantized theory, it has to further compensate a power divergence.
The heavy quark field $\heavy$ obeys $\frac{1+\gamma_0}{2}\heavy=\heavy$, 
and all spatial derivatives are symmetrized:
\begin{align}  \label{e:symdrv}
    \drvsym{i}  &= \tfrac{1}{2}( \drv{i} + \drvstar{i} ) \;, &
    \nabS{i}    &= \tfrac{1}{2}( \nab{i} + \nabstar{i} ) \;, &
   \nabSleft{i} &= \tfrac{1}{2}( \nableft{i} + \nableftstar{i} ) \;.
\end{align}

\begin{table}[t]
  \small
  \centering
  \renewcommand{\arraystretch}{1.2}
  %
%
\begin{tabular}{lcccccc}\toprule
     & \multicolumn{2}{c}{$\beta=5.5$} & \multicolumn{2}{c}{$\beta =5.3$} & \multicolumn{2}{c}{$\beta = 5.2$}  \\ 
       \cmidrule(lr){2-3}\cmidrule(lr){4-5}\cmidrule(lr){6-7}
                       & HYP1         & HYP2         & HYP1         & HYP2         & HYP1         & HYP2          \\ \midrule
 $a\mhbare^{\rm stat}$ & $0.969(10)$  & $1.000(10)$  & $1.317(13)$  & $1.350(13)$  & $1.520(15)$  & $1.554(15)$  \\ 
   $-\ln(\zastat)$     & $0.271(5) $  & $0.181(5) $  & $0.283(5) $  & $0.177(5) $  & $0.291(6) $  & $0.177(6) $  \\ \cmidrule(lr){2-7}
   $am_{\rm bare}$     & $0.594(16)$  & $0.606(16)$  & $0.993(18)$  & $1.014(18)$  & $1.214(19)$  & $1.239(19)$  \\ 
   $-\ln(\zahqet)$     & $0.156(42)$  & $0.163(36)$  & $0.169(37)$  & $0.146(32)$  & $0.169(35)$  & $0.136(31)$  \\ 
   $-\cahqet/a$        & $0.07(12) $  & $0.67(12) $  & $0.00(10) $  & $0.55(10) $  & $0.01(9)  $  & $0.54(9)  $  \\ 
   $\omegakin/a$       & $0.520(13)$  & $0.525(13)$  & $0.415(10)$  & $0.419(10)$  & $0.378(9) $  & $0.380(9) $  \\ 
   $\omegaspin/a$      & $0.949(40)$  & $1.090(46)$  & $0.731(31)$  & $0.883(37)$  & $0.655(27)$  & $0.812(33)$  \\ 
\bottomrule
\end{tabular}

  \caption{HQET parameters at the physical point 
           $\boldsymbol{\omega}(z=\zb)$.
            The parameters are given for $\zb$ determined 
            such that $\mB=5279.5\,\MeV$~\cite{Bernardoni:2013xba}, 
            which corresponds to $\zb^{\rm stat}=13.24(25)$ at static order, 
            and to $\zb=13.25(26)$ for HQET expanded to $\Or(\minvh)$.
            The bare coupling is $g_0^2=6/\beta$.}
  \label{tab:omega_zb}
\end{table}

In QCD the decay constant $\fBq$ is given by $\langle 0|\bar{q}\gamma_0
\gamma_5 b |B_q(\vecp=0)\rangle = \fBq \mBq$, with the relativistic convention
$\langle B_q(\vecp^\prime)|B_q(\vecp)\rangle = 2E_q(p)\delta(\vecp-\vecp^\prime)$.
We are thus interested in extracting matrix elements from correlation functions
defined at zero spatial momentum; the operator $A_0^{(2)}$ therefore does not
enter into our present computations at all.

In order to assure the renormalizability of HQET at next-to-leading order, the
$\Or(\minvh)$ terms in \eqref{e:hqetaction} are treated in the usual way as
operator insertions into static correlation functions,
\begin{equation}
  \langle O\rangle = \langle O\rangle_{\rm stat} +
    \omegakin \, a^4{\sum}_x \langle O\,\Okin(x) \rangle_\stat +
    \omegaspin\, a^4{\sum}_x \langle O\,\Ospin(x)\rangle_\stat  \;,
\end{equation}
where the suffix ``stat'' reminds us that expectation values are computed in
the static theory.

In our strategy we can treat HQET non-perturbatively to leading order (static)
or next-to-leading order, including terms of $\Or(\minvh)$. The corresponding
sets of parameters, $\boldsymbol{\omega}^\stat \equiv \big( \mhbare^\stat,
\zastat \big)$ or $\boldsymbol{\omega}\equiv \big( \mhbare, \omegakin,
\omegaspin, \zahqet, \cah{1} \big)$ absorb power and logarithmic divergences of
the effective theory regularized on the lattice. For technical reasons they
have been determined in
\cite{Blossier:2012qu}
for a series of heavy quark masses parameterized in terms of the
renormalization group invariant (RGI) heavy quark mass $z\equiv M L_1$.%
\footnote{The length scale $L_1$ is implicitly defined through the renormalized
coupling in the Schr\"odinger Functional scheme $\gbsq_{\rm SF}(L_1/2)=2.989$,
see~\cite{Blossier:2012qu,Fritzsch:2012wq}.}%
After our recent determination of the RGI b-quark mass $\zb\equiv \Mb
L_1$~\cite{Bernardoni:2013xba}, we now choose a quadratic polynomial to
interpolate $\boldsymbol{\omega}(z)$ --- computed at $z=11, 13, 15$ --- to the
physical point $\zb=13.25$. Similarly, we interpolate
$\boldsymbol{\omega}^\stat(z)$ to $\zb^\stat=13.24$ at the static order. As
expected from~\cite{Blossier:2012qu}, all individual interpolations of the
$\boldsymbol{\omega}^{(\stat)}(z)$ parameters to $\zb^{(\stat)}$ are smooth and
do not deviate much from the closest point at $z=13$.  In the following we will
refer to the HQET parameters at the physical b-quark mass only. For
completeness they are collected in Table~\ref{tab:omega_zb} for the three
lattice spacings $a(\beta)$ and two static discretizations (HYP1, HYP2) in use.
%
\subsection{Isolating the ground state}

\begin{figure}[t]
  \centering
  \includegraphics[width=\textwidth]{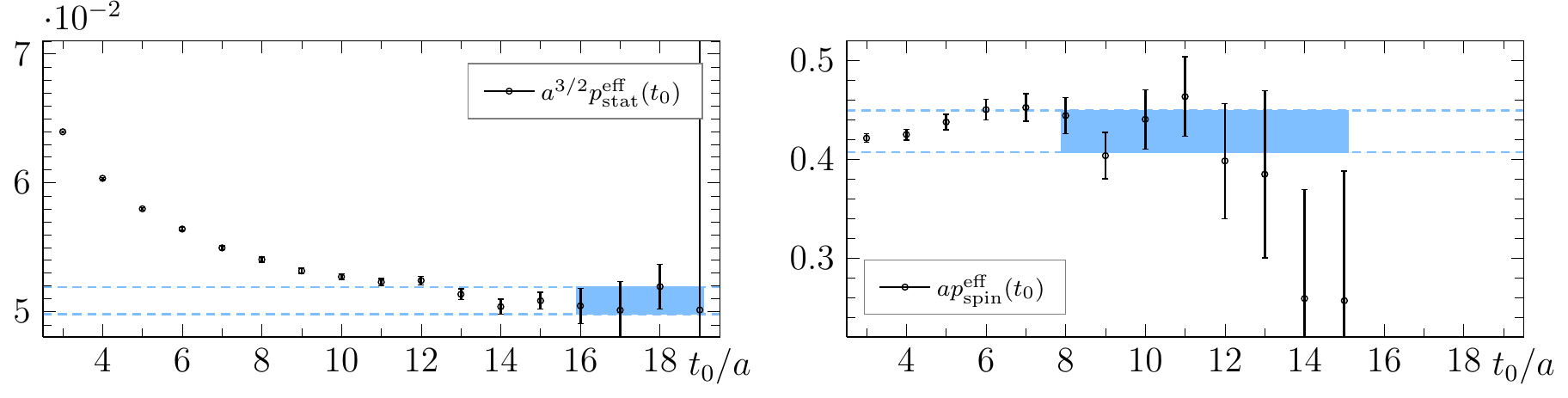}\\[0.2em]
  \includegraphics[width=\textwidth]{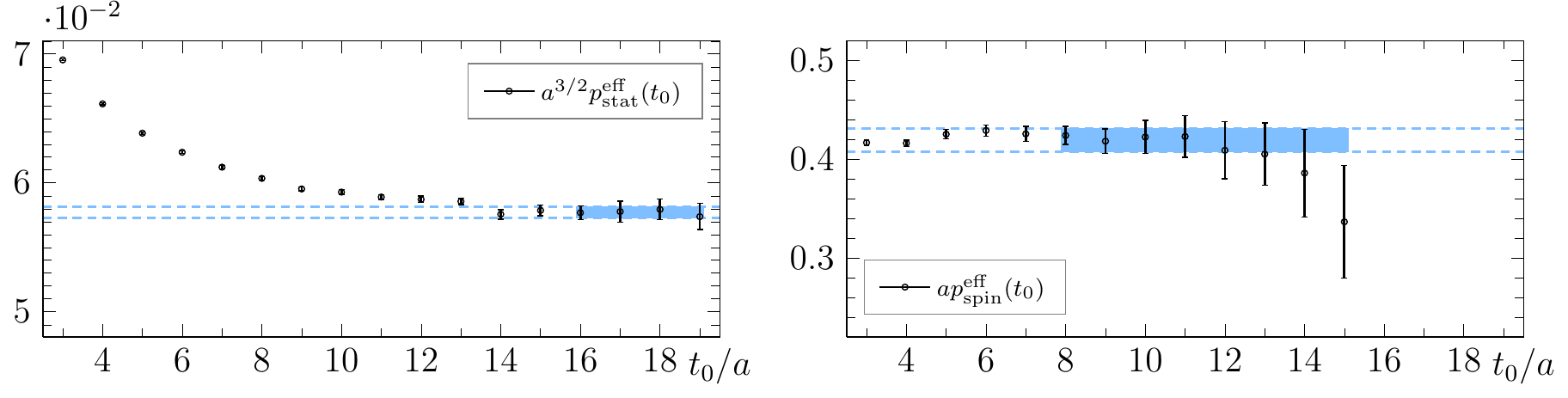}\\[-0.8em]
  \caption{Typical plateau averages after applying the GEVP analysis to data
           obtained on the $\Nf=2$ CLS ensemble N6 ($a=0.048\,\fm$, 
           $\mpi=340\,\MeV$). The two plots on top show our result
           for the B-meson static matrix element $p^{\rm stat}$ (left panel)
           and the $\Or(\minvh)$ chromomagnetic matrix element $p^{\rm spin}$ 
           (right panel). In the lower plots we present the corresponding
           quantities for the $\Bs$-meson.
          }
  \label{fig:plat_N6}
\end{figure}

In extracting hadronic quantities, it is crucial to have good control over the
contributions of excited states to the correlators.  Since we are interested in
the lowest-lying state in a given channel, we attempt to suppress excited-state
contaminations by considering appropriate linear combinations of correlation
functions.  Specifically, we form the matrices 
\begin{eqnarray} 
\label{e:cmatdefs}
  C^\stat_{ij}(t) & = & 
    \sum_{x,\vecy}  \;\left< O_i(x_0+t,\vecy)\,O^*_j(x)\right>_\stat \,, \\[-0.5ex]\nonumber
  C^{\rm{kin/spin}}_{ij}(t) & = & 
    \sum_{x,\vecy,z}\left< O_i(x_0+t,\vecy)\,{\cal O}_{\rm kin/spin}(z)\, O^*_j(x)\, \right>_\stat \;,
\end{eqnarray}
which depend on a basis of operators $O_i\,,\;i=1,\ldots,N$.  The key step is
to find a solution to the generalized eigenvalue problem (GEVP) in the static
approximation
\begin{eqnarray} \label{e:gevp}
  C^{\rm stat}(t)\, v^{\rm stat}_n(t,t_0) = \lambda^{\rm stat}_n(t,t_0)\, C^{\rm stat}(t_0)\,v^{\rm stat}_n(t,t_0) \,,
  \quad n=1,\ldots,N\,,\quad t>t_0\;.
\end{eqnarray}
Then, exploiting the orthogonality property of the eigenvectors,
\begin{equation} \nonumber
  \langle v_m^\stat(t,t_0) C^\stat(t_0)  v_n^\stat(t,t_0)\rangle  \propto 
  \delta_{nm}\;,
\end{equation} 
one can show that the $\Or(\minvh)$ corrections to the matrix elements
$\langle 0|\psibar_{\rm h}\gamma_0\gamma_5 \psi_q|B^{(n)}\rangle$ only depend
on the static generalized eigenvalues $\lambda_n^\stat(t,t_0)$, the vectors
$v_n^\stat(t,t_0)$, and the $\Or(\minvh)$ correlators $C^{\rm kin/spin}(t)$
\cite{Blossier:2009kd},
in analogy with perturbation theory in quantum mechanics.

We define the operator basis
\begin{align}
 O_k(x) &= \heavyb(x)\,\gamma_0\gamma_5\,\psi_q^{(k)}(x)            \,,&
  \psi_q^{(k)}(x) &= \left( 1+\kappa_{\rm G}\,a^2\,\Delta \right)^{R_k} \psi_q(x) \,,
\end{align}
where $\psi_{\rm h}(x)$ is the static quark field and $\psi_q^{(k)}(x)$
($q={\rm u/d}$ or $q={\rm s}$) is a Gaussian smeared~\cite{Gusken:1989ad}
relativistic quark field.  The gauge links in the covariant Laplacian $\Delta$
have been triply APE smeared~\cite{smear:ape,Basak:2005gi} in the spatial
directions.

The  parameters $\kappa_{\rm G}$ and $R_k$ have been chosen such that they
correspond to approximately the same sequence of physical radii at each value
of the lattice spacing, see~\cite{Bernardoni:2013xba} for details.  We solve
the GEVP for the matrix of correlators in the static limit, eq.~\eqref{e:gevp}
for $N=3$.  The resulting eigenvalues and eigenvectors, together with the
matrices in eq.~\eqref{e:cmatdefs} and the correlators
\begin{eqnarray}
   C^{\rm{stat/(1)}}_{A_0,j}(t)& = &
   \sum_{x,\vecy}
   \left< A^{\rm stat/(1)}_0(x_0+t,\vecy) O^*_j(x)\right>_\stat \,,
\end{eqnarray}
are used to build optimal interpolating fields such that the matrix elements
$\langle 0 | A_0^{\rm stat}| B^{(n)} \rangle$ and their $\Or(\minvh)$
corrections can be extracted, from the correlation functions above, up to
contaminations from excited states which are 
$\Or(e^{-(E^{\rm stat}_{N+1}-E^{\rm stat}_{n})t_{0}})$.  
Notice that for the ground state the energy difference in the exponential
correction is of the form $E_{N+1}-E_1$ rather than $E_{2}-E_1$,  with $N$ the
rank of the correlator matrices in eq.~\eqref{e:cmatdefs}.  This asymptotic
convergence holds for $t_{0} \geq t/2$, as discussed in detail
in~\cite{Blossier:2009kd}, to which we refer for any unexplained notation.  In
particular, the symbols $p^{\rm stat/x/A^{(1)}}=p^{\rm stat/x/A^{(1)}}_{n=1}$,
with x=kin/spin, are used in the following to indicate the static and
$\Or(\minvh)$ contributions to the matrix elements of the axial current as
in~\cite{Blossier:2010mk}, where one can find expressions for the expected time
dependence of the different terms, which read
\begin{eqnarray}
\nonumber
p^{\rm eff,\,stat}_{n}(t,t_{0})&=&p^{\rm stat}_{n}+\gamma^{\rm stat}_{n,N}
e^{-(E^{\rm stat}_{N+1}-E^{\rm stat}_{n})t_{0}}\;,\\
\nonumber
p^{\rm eff,\,x}_{n}(t,t_{0})&=&p^{\rm x}_{n}+\left[\gamma^{\rm x}_{n,N}
-\frac{\gamma^{\rm x}_{n,N}}
{p^{\rm stat}_{n}}t_{0}(E^{\rm x}_{N+1}-E^{\rm x}_{n})\right]
e^{-(E^{\rm stat}_{N+1}-E^{\rm stat}_{n})t_{0}}\;,\\
p^{\rm eff,\,A^{(1)}}_{n}(t,t_{0})&=&p^{\rm A^{(1)}}_{n}+\gamma^{\rm A^{(1)}}_{n,N}
e^{-(E^{\rm stat}_{N+1}-E^{\rm stat}_{n})t_{0}}\;.
\label{GEVPexpr}
\end{eqnarray}
We provide in Figure~\ref{fig:plat_N6} an illustration of typical plateaux for
the heavy-light and heavy-strange mesons matrix elements $p^\stat$ and $p^{\rm
spin}$. Those plateaux are chosen following the procedure detailed
in~\cite{Blossier:2010vz,Blossier:2010mk}.  The criteria use the results of the
GEVP analysis to ensure that in the plateau region the systematic errors due to
excited-state contributions are less than a given fraction (typically one
third) of the statistical errors.  As a consistency check, we have also
employed a global fit of the form of eqs.~\eqref{GEVPexpr} to our data. The
values of $p_n$ obtained from the fit were consistent with the plateau values,
albeit with smaller statistical errors. Our errors may therefore be seen as a
conservative estimate.

\subsection{B-meson decay constants at different orders in HQET}

The quantities of interest are obtained by combining the lattice parameters of
HQET, computed non-perturbatively, and the bare matrix elements evaluated in
the static theory.  All divergences of the effective quantum field theory are
thus properly removed and the continuum limit can be safely taken at a fixed
order in the $\minvh$ expansion.
\begin{table}
  \small
  \centering
  %
%
\begin{tabular}{clllllll} \toprule
       &     & \multicolumn{2}{c}{$\fB$\,[\MeV]}& \multicolumn{2}{c}{$\fBs$\,[\MeV]}  & \multicolumn{2}{c}{$\fBs/\fB$} \\ 
       \cmidrule(r){3-4}\cmidrule(r){5-6}\cmidrule(r){7-8}
 $e$-id& $y$        & HYP1     & HYP2     & HYP1     & HYP2     & HYP1      & HYP2       \\ \midrule
    A4 & 0.0771(14) & 212(9)   &  210(10) &  227(8)  &  227(8)  & 1.071(28) & 1.084(23)  \\
    A5 & 0.0624(13) & 206(7)   &  204(7)  &  226(6)  &  224(6)  & 1.096(20) & 1.100(19)  \\
    B6 & 0.0484(9)  & 198(8)   &  195(7)  &  224(8)  &  223(7)  & 1.127(36) & 1.144(32)  \\[0.15em]
    E5 & 0.0926(15) & 215(7)   &  213(8)  &  232(8)  &  231(9)  & 1.077(28) & 1.086(25)  \\
    F6 & 0.0562(9)  & 203(8)   &  201(8)  &  228(7)  &  228(7)  & 1.120(48) & 1.138(39)  \\
    F7 & 0.0449(7)  & 201(6)   &  200(6)  &  222(6)  &  223(7)  & 1.103(26) & 1.119(24)  \\
    G8 & 0.0260(5)  & 190(8)   &  190(8)  &    --    &  --      &    --     & --         \\[0.15em]
    N5 & 0.0940(24) & 222(16)  &  221(15) &    --    &  --      &    --     & --         \\
    N6 & 0.0662(10) & 205(14)  &  205(15) &  229(15) &  231(15) & 1.115(50) & 1.126(46)  \\
    O7 & 0.0447(7)  & 199(14)  &  194(14) &    --    &  228(14) &    --     & 1.178(85)  \\\midrule
    LO & $y^{\rm exp},a=0$ & \multicolumn{2}{c}{188(12)}
                                & \multicolumn{2}{c}{225(13)}  
                                & \multicolumn{2}{c}{1.184(60)}  \\
   NLO & $y^{\rm exp},a=0$ & \multicolumn{2}{c}{186(12)}
                                & \multicolumn{2}{c}{--}  
                                & \multicolumn{2}{c}{1.203(61)}  \\
    \bottomrule
\end{tabular}

  \caption{Raw data for $\fB$, $\fBs$ and their ratio $\fBs/\fB$, 
  using HQET parameters at the physical point $\boldsymbol{\omega}(z=\zb)$,
  with $\zb=13.25$ as determined in~\cite{Bernardoni:2013xba}. The last two
  rows summarize our results of a combined chiral and continuum extrapolation
  using either the LO or the NLO fit ansatz~\eqref{eq:fBextr} for each
  individual observable. 
  }
  \label{tab:fBq_full}
\end{table}

To obtain $\fB$ and $\fBs$ including $\Or(\minvh)$ terms in HQET, we compute
\begin{align}
  \phi_i &=  \ln(\zahqet) + \bastat a \mqi + \left.\left(\ln( a^{3/2}p^\stat)
              + \omegakin  \, p^{\rm kin}  
              + \omegaspin \, p^{\rm spin}
              + \cah{1}    \, p^{\rm A^{(1)}_0}\right)\right|_{m_{{\rm q},i}}  \,,\notag  \\
   \fBi  &= \exp( \phi_i )\big/\sqrt{a^3\,\mBi/2}   \,.\label{e:def-fB}  
\end{align}
Here, $i$ labels the light quark content (u/d- or s-quark) and the term
multiplying $\bastat$ is needed for the $\Or(a)$ improvement of mass-dependent
cutoff effects in the heavy-light axial current.%
\footnote{%
The bare mass $a\mqi=(1/\kappa_i-1/\hopc)/2$ is obtained from $\hopc$, the
point where the PCAC mass vanishes.
}%
The corresponding expression in static HQET is given by
\begin{align}
  \phi_i^\stat &=  \ln(\zastat) + \bastat a \mqi + \left.\left(\ln( a^{3/2}p^\stat)
                   +a\castat p^{\rm A^{(1)}_0} \right)\right|_{m_{{\rm q},i}}           \,,\notag \\
    \fBi^\stat &= \exp\{ \phi_i^\stat \}\big/\sqrt{a^3\,\mBi/2} \,,\label{e:def-fBstat}
\end{align}
where $\castat$ is another $\Or(a)$ improvement coefficient that is needed at
the  static order. Both $\bastat$ and $\castat$ have been computed
perturbatively in~\cite{Grimbach:2008uy}. If treated as an independent
observable, the ratio $\fBs/\fB$ is easily obtained through
eq.~\eqref{e:def-fB} at next-to-leading order, \viz
\begin{align}
     \fBs/\fB &= \exp\{ \phi_{\rm s} - \phi \}\big/\sqrt{\mBs/\mB}  \label{e:def-fBsfB} \,, 
\end{align}
or in the very same way through eq.~\eqref{e:def-fBstat} at the static order.
Note that in the ratio the leading dependence on the scale setting procedure,
which explicitly enters via the lattice spacing appearing as $a^{3/2}$,
cancels.  Furthermore, terms in eq.~\eqref{e:def-fB} or
eq.~\eqref{e:def-fBstat}, which do not carry an explicit label $i$ drop out and
the term multiplying $\bastat$ becomes independent of the critical hopping
parameter $\hopc(g_0^2)$.

\begin{table}[t]
  \small
  \centering
  %
%
\begin{tabular}{clllllll} \toprule
       &     & \multicolumn{2}{c}{$\fB^\stat$\,[\MeV]}& \multicolumn{2}{c}{$\fBs^\stat$\,[\MeV]}  & \multicolumn{2}{c}{$\fBs^\stat/\fB^\stat$} \\ 
       \cmidrule(r){3-4}\cmidrule(r){5-6}\cmidrule(r){7-8}
  $e$-id & $y$        & HYP1    & HYP2   & HYP1   & HYP2   & HYP1      & HYP2       \\ \midrule
    A4 & 0.0771(14) & 240(4)  & 228(4) & 264(5) & 250(4) & 1.101(9)  & 1.096(7)   \\
    A5 & 0.0624(13) & 235(4)  & 223(4) & 265(5) & 249(4) & 1.128(6)  & 1.117(5)   \\
    B6 & 0.0484(9)  & 224(5)  & 213(4) & 259(4) & 244(4) & 1.154(20) & 1.143(15)  \\[0.15em]
    E5 & 0.0926(15) & 240(4)  & 231(4) & 263(4) & 252(4) & 1.092(10) & 1.090(8)   \\
    F6 & 0.0562(9)  & 224(5)  & 214(4) & 257(4) & 245(4) & 1.149(18) & 1.148(16)  \\
    F7 & 0.0449(7)  & 219(4)  & 210(3) & 252(4) & 241(4) & 1.152(10) & 1.144(10)  \\
    G8 & 0.0260(5)  & 212(4)  & 205(4) &  --    &  --    &   --      &  --        \\[0.15em]
    N5 & 0.0940(24) & 241(6)  & 236(6) &  --    &  --    &   --      &  --        \\
    N6 & 0.0662(10) & 225(7)  & 217(5) & 254(4) & 245(4) & 1.129(24) & 1.133(18)  \\
    O7 & 0.0447(7)  & 217(9)  & 208(7) &  --    & 244(6) &    --     & 1.172(39)  \\\midrule
    LO & $y^{\rm exp},a=0$ & \multicolumn{2}{c}{192.5(52)}
                                & \multicolumn{2}{c}{234.1(48)}  
                                & \multicolumn{2}{c}{1.219(25)}  \\
   NLO & $y^{\rm exp},a=0$ & \multicolumn{2}{c}{190.3(51)}
                                & \multicolumn{2}{c}{--}  
                                & \multicolumn{2}{c}{1.189(24)}  \\
    \bottomrule
\end{tabular}

  \caption{Raw data for $\fB^\stat$, $\fBs^\stat$ and their
  ratio $\fBs^\stat/\fB^\stat$, 
  using static HQET parameters at the physical point
  $\boldsymbol{\omega}^\stat(z=\zb^\stat)$,
  with $\zb^\stat=13.24$ as determined in~\cite{Bernardoni:2013xba}. The last two
  rows summarize our results for a combined chiral and continuum extrapolation
  using either the LO or NLO fit ansatz~\eqref{eq:fBextr} for each individual 
  observable. 
  }
  \label{tab:fBqstat_full}
\end{table}

Concerning the parameters and overall statistics of the large-volume
simulations used in the present analysis we refer the reader to Table~1
of~\cite{Bernardoni:2013xba}.  The light quark is treated in a unitary setup
with $\mpi$ covering a range from $190\,\MeV$ to $440\,\MeV$ while the bare
strange quark mass is tuned on each CLS ensemble using the kaon decay constant
\cite{Fritzsch:2012wq}. The lattice spacings are
$a/\fm\in\{0.048,0.065,0.075\}$ for $\beta\in\{5.5,5.3,5.2\}$, corresponding to
the CLS ensemble ids N-O, E-G and A-B respectively. 
In Table~\ref{tab:fBq_full} we give our results for $\fB$, $\fBs$ and $\fBs/\fB$ 
as computed on these ensembles together with the results obtained after 
performing different chiral and continuum extrapolations to the physical point
$(\mpi,a)=(\mpi^{\rm exp},0)$. The latter are being discussed in more detail in
Section~\ref{sec:extrapolation}. Finally, we collect the values of the 
static quantities $\fB^\stat$, $\fBs^\stat$ and $\fBs^\stat/\fB^\stat$ in 
Table~\ref{tab:fBqstat_full}.

\subsection{Error analysis and propagation}

\begin{figure}[t]
   \centering
   \includegraphics[width=\textwidth]{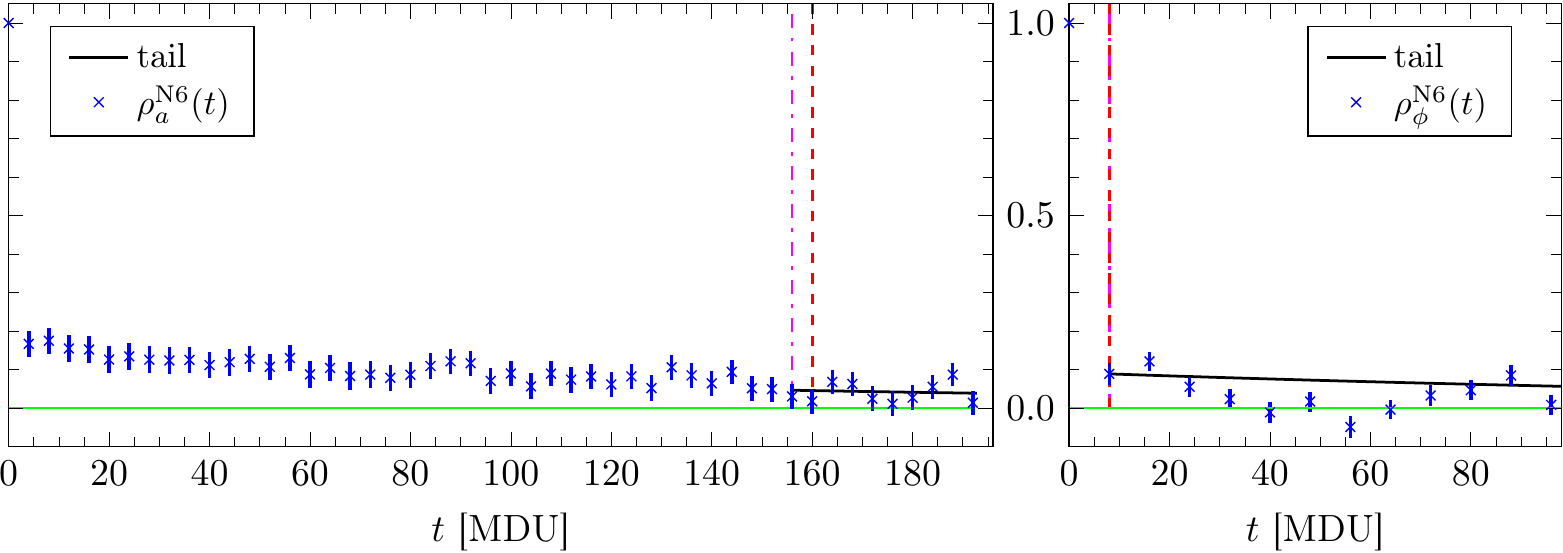}
   \vskip-0.8em
   \caption{Example of long-tail contributions to the total budget
            from ensemble N6 ($\tau_{\rm exp}^{\rm N6}=200$\,MDU). On the left we 
            plot the normalized autocorrelation function for the lattice
            spacing, $\rho^{\rm N6}_a(t)=\Gamma^{\rm N6}_a(t)/\Gamma^{\rm N6}_a(0)$,
            and on the right accordingly for the quantity $\phi$ defined in
            eq.~\eqref{e:def-fB}. For the lattice spacing data that enters
            $\rho_a$, measurements have been performed on each stored 
            configuration, separated by 4\,MDU, while for $\phi$ measurements
            are separated by 8\,MDU.}
   \label{fig:spacing-err-N6}
\end{figure}

We follow~\cite{Wolff:2003sm,Schaefer:2010hu} for all sources of errors. All
results or intermediate quantities are considered as functions $f(\pbar,Y)$ of
the means $\pbar_\alpha(\ens)=N_\ens^{-1}\sum_{m=1}^{N_\ens}p_\alpha^m(\ens)$
of primary MC data $p_\alpha^m(\ens)$ originating from configuration $m$ of the
ensemble number $\ens$ (corresponding to $\ens$-id in Table~\ref{tab:fBq_full}
and \ref{tab:fBqstat_full}), as well as functions of additional input $Y$, such
as the HQET parameters $\omega_i$.  Also the results of fits to the data are
considered as functions of the original data, where the weights in the fits (we
always use only the diagonal errors as weights) are precomputed and then kept
fixed, i.e., a dependence of $f$ on the weights is not considered. 

The error $\sigma_f$ of such a function is then
\begin{align}
     \sigma_f^2 &= \sum_\ens \sigma_f^2(\ens) 
   + \sum_{i,j} \frac{\partial f}{\partial{Y_i}} \, C^Y_{ij} \,
      \frac{\partial f}{\partial{Y_j}} \;.
\end{align}
The block-diagonal covariance matrix $C^Y$ of the additional input is known: a
block~\cite{Fritzsch:2012wq} for the axial current renormalization factors at
the three different $\beta$ (entering the lattice spacing determination and
$\fpi$) and a block~\cite{Blossier:2012qu} for the $\omega_i$. The
contributions from the individual ensembles $\ens$ are 
\begin{align}
  \sigma_f^2(\ens) &= \frac1{N_\ens} \left[ 
   \Gamma^\ens_{f}(0) + 
   2 \sum_{m=1}^{W-1}\Gamma^\ens_{f}(m) +
   2\tau_\mathrm{exp}^\ens \Gamma^\ens_{f}(W) \right]\;, \nonumber
   \\
   \Gamma^\ens_f(m) &= \sum_{\alpha,\beta} \frac{\partial f}{\partial \pbar_\alpha} 
       \Gamma^\ens_{\alpha\beta}(m) \frac{\partial f}{\partial \pbar_\beta}\;.
   \label{e:sigma_ens}     
\end{align}
The term proportional to $\tau^\ens_\mathrm{exp}$ accounts for the
difficult-to-estimate contribution of the tails to the autocorrelation function
$\Gamma_f^\ens$~\cite{Schaefer:2010hu}. For $\tau^\ens_\mathrm{exp}$ we insert
our previously estimated values (see e.g. Table 1
of~\cite{Bernardoni:2013xba}), and $W$ is chosen as the point where
$\Gamma^\ens_f$ comes close to zero within about (1-2 $\times$) its estimated
statistical error.  The required derivatives $\frac{\partial f}{\partial
\pbar_\alpha}$ are computed numerically~\cite{Wolff:2003sm}.

We note that there are many hidden correlations which are all taken into
account, e.g., the lattice spacing at one $\beta$ depends on information from
other $\beta$ through the combined chiral extrapolation
in~\cite{Fritzsch:2012wq}.  A straightforward implementation of
\eq{e:sigma_ens} would be cumbersome and numerically expensive. We  compute it
iteratively instead~\cite{ALPHAerr,ALPHAerr:LP14}.

As explicit example we choose the ensemble with the highest statistics
available, $\ens={\rm N6}$.~\Fig{fig:spacing-err-N6} shows the numerical
estimate of the normalized autocorrelation function $\rho(t)$ in terms of the
simulation time in molecular dynamic units (MDU). After summing up the
autocorrelation function explicitly within a window where it is still rather
well determined, the sum up to infinity is determined by modelling it with a
single exponential $\exp(-t/\tau_\mathrm{exp})$ plotted as ``tail''. 
On the left hand side of the figure the observable chosen is the lattice
spacing (see \cite{Fritzsch:2012wq}); the relevant contribution is from the
kaon decay constant. On the right hand side we have chosen $\phi$, i.e.,
essentially the B-meson decay constant in lattice units (see \eq{e:def-fB}).
While on the left, where only light-quark physics enters and measurements were
taken more frequently, the tail is seen quite well at $t < 100$~MDU, on the
right the autocorrelation function appears to drop to a lower value at short
time and in fact is not significant at $t \approx 30$~MDU. Still, our somewhat
conservative procedure estimates a $\approx 35\%$ contribution to
$\tau_\mathrm{int}$ on the left and $\approx 82\%$ on the right.

\begin{table}
  \small
  \centering
  %
%
\begin{tabular}{ccccccccc}\toprule
    $\beta=6/g_0^2$ & $\hopc(g_0^2)$  &\multicolumn{2}{c}{$\castat(g_0^2)$}&\multicolumn{2}{c}{$\bastat(g_0^2)$} \\\cmidrule(lr){3-4}\cmidrule(lr){5-6}
            &           &  HYP1      &  HYP2      &  HYP1      &  HYP2         \\\midrule
    5.2     & 0.1360546 &  0.0033461 &  0.0597692 &  0.6045384 &  0.6638461  \\
    5.3     & 0.1364572 &  0.0032830 &  0.0586415 &  0.6025660 &  0.6607547  \\
    5.5     & 0.1367749 &  0.0031636 &  0.0565090 &  0.5988363 &  0.6549090  \\
\bottomrule
\end{tabular}

  \caption{Numerical values of the improvement coefficients $\bastat$ and
           $\castat$ from 1-loop PT~\cite{Grimbach:2008uy}.}
  \label{tab:bAcAstat}
\end{table}

\section{Continuum and chiral limit extrapolations}\label{sec:extrapolation}

\begin{figure}
  \small 
  \centering\label{fig:extrfB}
  \includegraphics[width=\textwidth]{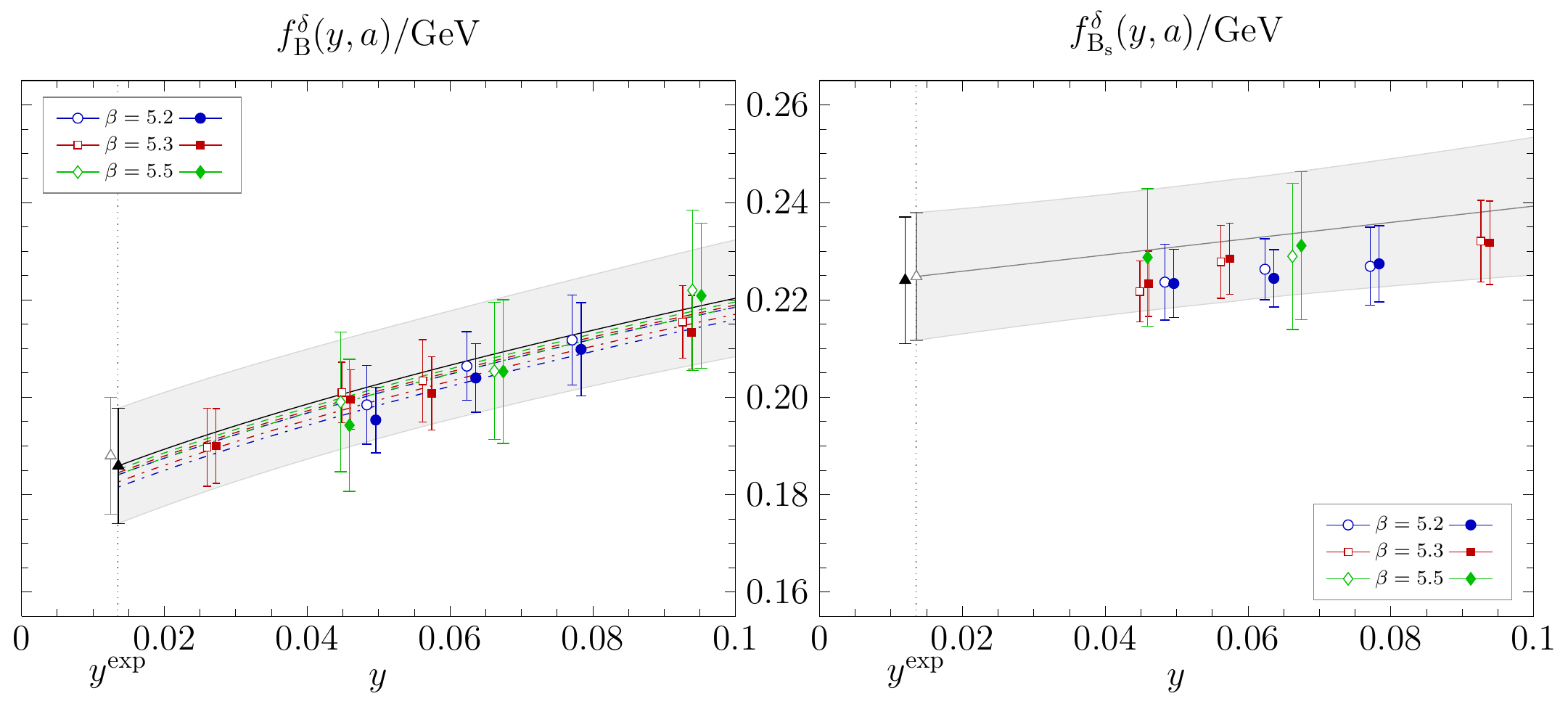}
  \vskip-0.8em
  \caption{Extrapolation of the $B$ (left panel) and $B_{\rm s}$ (right panel) 
           meson decay constant to the physical point. On the left, 
           the extrapolation using HM$\chi$PT at NLO (filled triangle) is 
           compared to a linear one (open triangle), in order to extract the systematic 
           error from truncating HM$\chi$PT at NLO. For $\fBs$ only a LO
           formula is known and shown. As a comparison we also add our final
           result, the continuum value of $\fBs=[\fBs/\fB]\fB$.
           All data points are listed in Table~\ref{tab:fBq_full}.}
\end{figure}

We use formulae from Heavy Meson Chiral Perturbation Theory (HM$\chi$PT) when
applicable \cite{Goity:1992tp, Sharpe:1995qp, Bernardoni:2009sx}: 
\begin{align} \notag
     \sqrt{\tfrac{\mB}{2}}  \fB^{\delta}(y,a)  
      &= A\left[1-\tfrac{3}{4}\tfrac{1 + 3 \widehat{g}^2}{2} \big\{ y\log( y ) - y^{\rm exp} \log( y^{\rm exp} ) \big\}\right] 
         + C\, (y - y^{\rm exp}) + D^{\delta}\, a^2 \;,\\  \label{eq:fBextr}
     \sqrt{\tfrac{\mBs}{2}} \fBs^{\delta}(y,a)
      &= A_{\rm s}+C_{\rm s}\,(y - y^{\rm exp}) + D_{\rm s}^\delta\, a^2\;.
\end{align}
As in~\cite{Bernardoni:2013xba} we have parameterized the chiral behaviour
through the variable $y=\mpi^2\big/8\pi^2\fpi^2$, with $y^{\rm exp}$
representing the physical value at $\fpi^{\rm exp}=130.4\,\MeV$ and $\mpi^{\rm
exp}=134.98\,\MeV$. Since we employ two static discretizations, we also need to
account for different cutoff effects, parameterized by $D^{\delta}$, with
$\delta=1,2$ corresponding to HYP$\delta$.  Due to the universality of the
continuum limit, the other coefficients do not depend upon $\delta$.  After
having fixed the HQET parameters at the physical b-quark mass, see
Table~\ref{tab:omega_zb}, we treat the B-meson masses as constants, which are
taken as $\mB= 5279.5\,\MeV$ and $\mBs=5366.3\,\MeV$ from the
PDG~\cite{Beringer:1900zz}. 
Together with all data points that enter the joint chiral and continuum
extrapolation, we list our results from different, independent fit ans\"atze
in~\Tab{tab:fBq_full} and~\ref{tab:fBqstat_full}.
For consistency, we decided to treat $\fBs$ as the dependent observable, to be
derived from our final results
\begin{align}\label{eq:HQET_res}
              \fB  &= 186(13)(2)_\chi \,\MeV \;, &    
  {\fBs}\big/{\fB} &= 1.203(62)(19)_\chi     \;.
\end{align}
The first, statistical, error as obtained from the NLO HM$\chi$PT fit ansatz
also includes the discrepancy to the static result, the uncertainty from the
HQET parameters and the lattice spacings.  We add a second, systematic, error
to account for the uncertainty in the chiral extrapolation. It is given by the
difference between the quoted value and its counterpart obtained by employing
the LO fit ansatz for the chiral extrapolation.  While we show only the NLO
extrapolation of $\fB$ in the left panel of~\Fig{fig:extrfB}, we also add the
continuum extrapolated value from the LO fit ansatz.  With all correlations
taken into account, our estimate for the ${\rm B}_{\rm s}$-meson decay constant
becomes
\begin{align}\label{eq:HQET_fBs} 
  \fBs &=   224(14)(2)_\chi \,\MeV\;.
\end{align}
In the right panel of~\Fig{fig:extrfB}, we contrast this result (filled
triangle) with an extrapolation of our $\fBs$ lattice data as if treated as an
independent observable, c.f.~\Tab{tab:fBq_full}.  We have also tried a
continuum extrapolation keeping a term linear in $a$ in the fit functions.  In
fact, we have not included $\Or(a)$ improvement terms in the HQET action and
current insertions at $\Or(\minvh)$.  These effects, formally of 
$\Or(a/m_{\rm h})$, are expected to be small, and within our error we do not 
observe any such dependence.

To get an insight on the convergence in $\minvh$, it is interesting to compare
our estimates at subleading order with those at static order of HQET.  By
applying the same fit formulae as in eqs.~\eqref{eq:fBextr}, we obtain
\begin{align}
                         \fB^{\rm stat}  &= 190(5)(2)_\chi \,\MeV\;, & 
  \frac{\fBs^{\rm stat}}{\fB^{\rm stat}} &= 1.189(24)(30)_\chi,      &    
                         \fBs^{\rm stat} &= 226(6)(9)_\chi   \,\MeV\;. 
        \label{eq:stat_res}
\end{align}
In~\Tab{tab:err-dist} we split the statistical error of our observables among
different sources. Obviously, the errors from the HQET parameters $\omega$ and
renormalization factor $\za$ which enters through the scale setting, largely
cancel in the ratio.  Although one looses precision, in general, due to the
increased variance in HQET observables compared to observables in the light
quark sector (such as $\mpi$ or $\fpi$), one is in the fortunate position that
the former couple less to the slow modes of the Monte Carlo chain, and
therefore their integrated autocorrelation times are smaller than for ``light''
quantities.

\begin{table}
  \centering\small
  \def\sep{\hphantom{2}}
\begin{tabular}{ccccccc}
\toprule
 Source  &   $\fBs$     &  $\fB$       & $\fBs/\fB$   & $\fBs^{\rm stat}$ & $\fB^{\rm stat}$ & $\fBs^{\rm stat}/\fB^{\rm stat}$  \\\midrule
   A3    & \sep0.20 \%  & \sep0.19 \%  & \sep0.00 \%  & \sep1.22 \%       & \sep1.10 \%      & \sep0.00 \% \\  
   A4    & \sep5.94 \%  & \sep9.36 \%  &    14.27 \%  & \sep8.06 \%       & \sep2.76 \%      &    14.36 \% \\  
   A5    & \sep1.17 \%  & \sep6.51 \%  & \sep7.37 \%  & \sep2.01 \%       & \sep0.91 \%      & \sep3.10 \% \\  
   B6    & \sep3.32 \%  & \sep2.99 \%  & \sep0.00 \%  & \sep2.70 \%       & \sep1.44 \%      & \sep0.26 \% \\  
   E5    & \sep1.15 \%  & \sep1.28 \%  & \sep0.21 \%  & \sep1.00 \%       & \sep0.95 \%      & \sep0.01 \% \\  
   F6    & \sep1.70 \%  & \sep2.21 \%  & \sep6.44 \%  & \sep1.85 \%       & \sep2.62 \%      & \sep9.65 \% \\  
   F7    &    15.41 \%  & \sep5.79 \%  &    37.01 \%  &    14.89 \%       & \sep3.02 \%      &    40.32 \% \\  
   G8    &    13.96 \%  &    12.81 \%  & \sep0.00 \%  &    15.36 \%       &    13.26 \%      & \sep0.00 \% \\  
   N5    & \sep5.91 \%  & \sep5.43 \%  & \sep0.00 \%  & \sep9.17 \%       & \sep7.94 \%      & \sep0.00 \% \\  
   N6    &    19.42 \%  &    13.78 \%  &    29.87 \%  & \sep8.35 \%       &    24.10 \%      &    28.61 \% \\  
   O7    &    16.03 \%  &    25.46 \%  & \sep4.80 \%  &    19.91 \%       &    27.66 \%      & \sep3.58 \% \\  
$\omega$ &    14.02 \%  &    12.72 \%  & \sep0.01 \%  & \sep8.35 \%       & \sep7.21 \%      & \sep0.00 \% \\  
$\za$    & \sep1.77 \%  & \sep1.46 \%  & \sep0.01 \%  & \sep7.13 \%       & \sep7.04 \%      & \sep0.09 \% \\  
  \bottomrule
  \end{tabular}

  \caption{Distribution of relative squared errors among different sources
           for~\eqref{eq:HQET_res}--\eqref{eq:stat_res}.}
  \label{tab:err-dist}
\end{table}

\subsection{A quick look at phenomenology}

The Flavour Lattice Averaging Group (FLAG)~\cite{Aoki:2013ldr} has made a
selection of lattice results for $\fB$, $\fBs$ and $\fBs/\fB$ with $\Nf=2$,
$2+1$ and $2+1+1$ dynamical quarks~%
\cite{Dimopoulos:2011gx,Carrasco:2013zta,McNeile:2011ng,Bazavov:2011aa,Na:2012kp, Dowdall:2013tga}.
Only one determination entered the two-flavour average and has been
updated~\cite{Carrasco:2013zta} since. Their values $\fB=189(8)\,\MeV$,
$\fBs=228(8)$ and $\fBs/\fB=1.206(24)$ are fully compatible with ours.
Averaging both $\Nf=2$ results produces numbers which are consistent with the
estimate from $N_{\rm f}=2+1$ computations quoted by FLAG:
$\fB^{\Nf=2+1}=190.5(4.2)\,\MeV$, $\fB^{\Nf=2+1}=227.7(4.5)$ and
$\fBs/\fB=1.202(22)$.

\noindent As a phenomenological application, we can insert our results for
$\fB$ and $\fBs$ into the formulae describing the branching ratios of $B \to
\tau \nu_\tau$ and $B_{\rm s} \to \mu^+\mu^-$ transitions:
\begin{align}\label{eq:BRs}
   \mcB(B^- \to \tau^- \bar{\nu}_\tau) &= 
   \frac{G^2_{\rm F} |V_{\rm ub}|^2}{8\pi} \tau_{\rm B}^{\phantom{1}} \fB^2 \mB^{\phantom{1}}  m^2_\tau\times \left(1-\frac{m^2_\tau}{\mB^2}\right)^2 \;, \\\notag
   \mcB(B_{\rm s} \to \mu^+ \mu^-) &= 
   \frac{G^2_{\rm F}}{\pi}\left[\frac{\alpha_{em}(m_Z)}{4\pi \sin^2 \theta_{\rm W}}\right]^2
   \tau_{\rm B_s}^{\phantom{1}} \fBs^2 \mBs^{\phantom{1}} m^2_\mu \sqrt{1-\frac{4m^2_\mu}{\mBs^2}} \,
   |V^*_{\rm tb}V_{\rm ts}^{\phantom{1}}|^2 \, Y^2  \;.
\end{align}
Here $Y\equiv Y(x_{\rm tW}, x_{\rm Ht},\alpha_{s})$ takes into account various
electroweak and QCD corrections, parameterized by 
$x_{\rm tW}=m^2_{\rm t}/m^2_{\rm W}$ and $x_{\rm Ht}=M^2_{\rm H}/m^2_{\rm t}$ 
with $M_{\rm H}$ being the Higgs boson mass. Using as inputs the experimental 
value $\mcB(B \to \tau \nu_\tau)_{\rm exp}=1.05(25) \times 10^{-4}$ quoted by 
the PDG~\cite{Beringer:1900zz,Aubert:2007xj,Aubert:2009wt,Hara:2010dk,Adachi:2012mm} 
and our estimate of $\fB$, we get 
\begin{align}\label{eq:Vub}
   |V_{\rm ub}| &= 4.15\,(29)_{\fB}(48)_{\mcB} \times 10^{-3},
\end{align}
where the errors come from $\fB$ and the branching ratio, respectively. The
value is roughly 1.5 $\sigma$ above the exclusive determination from 
$B \to \pi \ell \nu$.

Moreover, using the recent combination of experimental measurements at LHC,
namely $\mcB(B_{\rm s} \to \mu^+ \mu^-)=(2.9 \pm 0.7 ) \times 10^{-9}$~
\cite{Aaij:2013aka,Chatrchyan:2013bka,CMS-PAS-BPH-13-007}, 
together with our determination of $\fBs$, and all input parameters
of~\eqref{eq:BRs} set as in~\cite{Buras:2012ru}, we obtain 
\begin{align}\label{eq:VtbVts}
   |V^*_{\rm tb} V_{\rm ts}^{\phantom{1}}| &= 3.89 \, (24)_{\fBs}(47)_{\mcB} \times 10^{-2} \,.
\end{align}
The number is in good agreement with the extraction from global fits, which is
mostly constrained by $B^0_{\rm s} - \overline{B^0_{\rm s}}$ mixing.

\section{Conclusions}\label{sec:conclusions}

\noindent In this paper we have reported on our lattice measurement of the
decay constants $\fB$ and $\fBs$ performed with two dynamical flavours of
$\Or(a)$ improved Wilson fermions. The b-quark is treated in HQET, with the
matching to QCD performed non-perturbatively.  This makes the computation
entirely non-perturbative, with no reference to continuum renormalized
perturbation theory at any point.  After an extrapolation to the chiral and
continuum limit, we obtain
\begin{align}\label{eq:res-final}
 \fB  &= 186(13)\,\MeV \;, &    
  {\fBs}\big/{\fB} &= 1.203(65) \;, & \fBs &=   224(14) \MeV\;. 
\end{align}
Though it is important to check the dependence of these results on the number
of dynamical flavours, and therefore to repeat the computation with a dynamical
strange quark,  it may still be interesting to compute the ratios 
$f_{\rm B^*}/\fB$ and $f_{\rm B^*_0}/\fB$ on the $\Nf=2$ ensembles.  The first 
one is often used to check the reliability of sum rules in the
B-sector~\cite{Bekavac:2009zc}. A lattice measurement at $\Or(1/\mb)$ requires
the matching coefficients that are being computed by the ALPHA Collaboration to
extract $B \to \pi \ell \nu$ form factors~\cite{DellaMorte:2013ega}.  The
second ratio, already in the static limit, can be used to gain some insight
into the precision of phenomenological applications of HM$\chi$PT, in
particular concerning the relevance of the contributions from the $J^P =
\{0^+,\,1^+\}$ doublet states in chiral loops~\cite{Becirevic:2004uv}. 

The method of the present paper to compute B-meson decay constants has been
used previously in the framework of quenched QCD to estimate $\fBs$ without
inclusion of virtual quark loops~\cite{Blossier:2010mk}.  There, the scale
$r_0$ defined via the static quark potential~\cite{Sommer:1993ce} was employed
to express the decay constant in physical units, corresponding to
$\fBs^{\Nf=0}=216(5)\,\MeV$ for $r_0=0.5\,\fm$ and $\fBs^{\Nf=0}=252(7)\,\MeV$
for $r_0=0.45\,\fm$.  Given the rather reliable evidence that the true $r_0$ in
physical units lies in between these values (see~\cite{Sommer:2014mea} for a
review of the current status), our final result in eq.~\eqref{eq:res-final} is
compatible with the quenched one at the present level of precision.  Hence, no
significant $\Nf$-dependence can be stated.

An interesting piece of information is also contained in the technical
\Tab{tab:err-dist}. It shows that the uncertainties in the non-perturbatively
determined HQET parameters contribute only at the level of 8\% in the static
limit and 14\% when $\minv$ terms are included.  Moreover, we find the
$\Or(\minv)$ corrections to be very small, $\lesssim 2.5\%$.  This, together
with the fact that the computation of the $\omega_i$ can be much improved with
today's machines, gives us confidence that errors can be significantly reduced
in the future computation with 2+1 dynamical flavours.

\begin{acknowledgement}%
We would like to thank S.~Lottini for many helpful discussions as well as for 
providing us with the latest results on $\fpi, \mpi$ and $\fK$ prior to their 
final publication. Furthermore, we appreciate the support of many colleagues 
within the CLS effort for the joint production and use of gauge configurations.
This work is supported in part by the SFB/TR~9, by grant HE~4517/2-1 (P.F. and
J.H.) and HE~4517/3-1 (J.H.) of the Deutsche Forschungsgemeinschaft, and by
the European Community through EU Contract MRTN-CT-2006-035482, ``FLAVIAnet''.
It was also partially supported by the Spanish Ministry of Education and
Science projects RyC-2011-08557 (M.D.M.). 
We gratefully acknowledge the computer resources
granted by the John von Neumann Institute for Computing (NIC)
and provided on the supercomputer JUROPA at J\"ulich
Supercomputing Centre (JSC) and by the Gauss Centre for
Supercomputing (GCS) through the NIC on the GCS share
of the supercomputer JUQUEEN at JSC,
with funding by the German Federal Ministry of Education and Research
(BMBF) and the German State Ministries for Research
of Baden-W\"urttemberg (MWK), Bayern (StMWFK) and
Nordrhein-Westfalen (MIWF), as well as
within the Distributed European Computing Initiative by the 
PRACE-2IP, with funding from the European Community's Seventh 
Framework Programme (FP7/2007-2013) under grant agreement RI-283493,
by the Grand \'Equipement National de Calcul Intensif at CINES in 
Montpellier under the allocation 2012-056808,
by the HLRN in Berlin, and by NIC at DESY, Zeuthen.
\end{acknowledgement}

\end{document}